\documentstyle[preprint,eqsecnum,aps]{revtex}
\begin{document}
\draft
\preprint{KYUSHU-HET-35}
\title{
  Perturbative Tamm-Dancoff Renormalization
}
\author{
    Koji Harada and
    Atsushi Okazaki
}
\address{
    Department of Physics, Kyushu University\\
    Fukuoka 812-81, Japan
}

\date{\today}
\maketitle
\begin{abstract}
    A new two-step renormalization procedure is proposed. In the
first
	step, the effects of high-energy states are considered in the
	conventional (Feynman) perturbation theory. In the
	second step, the coupling to many-body states is eliminated by a
	similarity transformation. The resultant effective Hamiltonian
	contains only interactions which do not change particle number.
	It is subject to numerical diagonalization.
	 We apply the general procedure to a simple example
	 for the purpose of illustration.
\end{abstract}

\narrowtext
\section{Introduction}
One of the most serious issues in Light-Front Field Theory (LFFT) is
the complicated renormalization procedure\cite{reviews}.
If we wish to solve
relativistic bound state problems by diagonalizing {\it finite}
dimensional
Hamiltonian matrices numerically, we are forced to make a kind of
Tamm-Dancoff (TD) truncation\cite{TD}, which restricts the total
number of
particles in any intermediate states. Although this has been very
powerful in two-dimensional field theories such as the massive
Schwinger model\cite{MSM}, it causes a serious problem in
the renormalization program in higher dimensions.
It is because loop diagrams contain more
particles in the intermediate states and thus may not be included in
the high Fock sectors in the TD approximation.
On the other hand, the corresponding
counterterms are not excluded by the TD truncation in a
similar way. This is the source of the so-called ``sector-dependent''
counterterms\cite{PHW}.

How is it possible to solve the relativistic bound state problem with
a {\it finite} dimensional Hamiltonian without hitting the
above-mentioned problem?
A field theoretical Hamiltonian is infinite dimensional in two
aspects:
(1) In a relativistic field theory, fluctuations of
all scales couple each other. A high-energy fluctuation (or state)
couples to a low-energy one in a similar way as a low-energy one
does. The Hamiltonian is therefore infinite in ``energy space.''
(2) A relativistic field theoretical Hamiltonian usually
contains interactions
which change the particle number of a state. If we
represent the Hamiltonian in a Fock space, many-body states
couple to few-body states. The Hamiltonian is infinite in ``particle
number space.''
In order to obtain an effective finite dimensional Hamiltonian matrix
we have to have the control over the high-energy states
and the many-body states.

Our hope is to simulate the effects of the many-body,
high-energy states  on the low-energy, few-body states by
a set of effective interactions which do not change particle number
so much and do not couple to high-energy states either.
In this way, we might be able to justify the description of a bound
state as a weakly bound system of the constituents.
The ``parton picture'' of bound states
is most apparent in light-front quantization.
This is precisely the reason why we are
interested in LFFT.

A renormalization group approach is necessary to control the effects
of high-energy and many-body states. The conventional notion of
renormalization is to simulate the effects of high-energy states by a
set of local operators, called counterterms. It is obvious from the
discussions in the previous paragraphs that
we need to generalize it so that counterterms also simulate the
effects of many-body states. It is useful to distinguish two
different
renormalizations, though they are closely related in
a complicated way.
(1) ``Energy renormalization'' controls the effects of high-energy
states. It is almost identical to the conventional one. The only
difference comes from light-front quantization; in equal-time theory,
a high-energy state is a high-momentum states, while on the light
front, a high-energy state may have a small momentum.
This implies that
the counterterms in LFFT are no longer local. Nevertheless we still
hope that we may use
the covariant perturbation theory in this renormalization.
(2) ``TD renormalization''\cite{foot} deals with the effects of
many-body states. Probably, the most useful scheme for the
TD renormalization is the
similarity transformation of the Hamiltonian\cite{GW,wegner}.
A similarity transformation does
not change the eigenvalues of the Hamiltonian. By appropriately
choosing the similarity transformation, one may make the Hamiltonian
more diagonal. The idea is to use similarity transformations to make
the Hamiltonian more diagonal {\it in particle number space}.

There are two advantages in this approach:
(1) We may be able to use
the covariant perturbation theory in the first step. This is
important because the covariant perturbation theory is very powerful
in revealing the divergence structure.
Furthermore, it is much easier to obtain
the effects of loops than in the old-fashioned perturbation theory,
or,
by a perturbative similarity transformation in energy space.
(2) The troubles of
TD truncations are avoided at least partially. The effects of
the higher Fock states are included as the effective (induced)
interactions generated in
the TD renormalization.

The similarity renormalization group approach is proposed by
G{\l}azek
and Wilson\cite{GW} (and by Wegner\cite{wegner} independently) and
discussed by the Ohio-State group extensively\cite{ohio}.
The similarity
renormalization group is designed to avoid small energy denominators
in the old-fashioned perturbation theory and to eliminate the
interactions which change the {\it energies} of the unperturbed
states
drastically. There are two problems in their approach:
(1) It is hard
to incorporate with loop diagrams. The coupling constants
do not run in the calculations so far done\cite{ohio}. As far as
they do not
run, it is easy to obtain the so-called ``coherent''
Hamiltonian\cite{PW}. On the other hand, however, if one goes
to higher
orders and takes into account the running, one would face to the
difficulty of getting it.
(2) After the similarity
transformation, there are still interactions which change the number
of particles. In the calculations so far done, such interactions are
usually ignored. In our two-step approach, we
take advantage of the covariant perturbation theory to
incorporate with
the loop contributions in the first stage, while the interactions
which change the number of particles are eliminated systematically.
The TD truncation now becomes a part of the approximation, which
can be improved systematically.

In this paper, we discuss a simple example to
illustrate the basic idea of the two-step perturbative TD
renormalization.
The model is a $(1+1)$-dimensional (discretized) field theory
which is a kind of
hybrid of $(1+1)$-dimensional equal-time $\phi^{4}$ theory and the
one
quantized on the light front. This model exhibits the mass, coupling
constant, and wave function renormalization at the second-order in
perturbation theory. We first obtain the counterterms in the
Feynman perturbation theory.
We then diagonalize the Hamiltonian in a naive Tamm-Dancoff
approximation. We see that the lowest energy state is almost
independent of the cutoff if the counterterms are included, while the
first excited state become more cutoff dependent if the
counterterms are included. This illustrates the necessity of the
sector-dependent counterterms in a naive Tamm-Dancoff approximation.

Next we eliminate the interactions which change the number of
particles perturbatively by a similarity transformation.
The resulting effective Hamiltonian is diagonal in particle
number space. We then diagonalize the effective Hamiltonian
numerically. The lowest energy state is almost independent of the
cutoff. It is one of our main results to show
that the next-to-lowest energy state
is also almost independent of the cutoff, despite that we work in a
severely truncated space.

\section{The model}
Let us begin with the definition of the model, which we call
``$a^{4}$ theory.'' The Hamiltonian is given by
\begin{eqnarray}
	H & = & \sum_{n=-\Lambda}^{\Lambda}\omega_{n}a_{n}^{\dagger}a_{n}
	+{g\over 6}\sum_{k,l,m,n=-\Lambda}^{\Lambda}\delta_{k,l+m+n}
	\left(a^{\dagger}_{k}a_{l}a_{m}a_{n}
	+a^{\dagger}_{n}a^{\dagger}_{m}a^{\dagger}_{l}a_{k}\right)
	\nonumber  \\
	 &  & {}+{\lambda\over 4}
	 \sum_{k,l,m,n=-\Lambda}^{\Lambda}\delta_{k+l,m+n}
	 a^{\dagger}_{k}a^{\dagger}_{l}a_{m}a_{n}
	\label{ham}
\end{eqnarray}
where ``momenta'' $k,l,m,n$ are integers.
The unperturbed energy $\omega_{n}$ is given
by
\begin{equation}
	\omega_{n}=|n|+\mu
	\label{omega}
\end{equation}
where $\mu$ may be interpreted as mass. We assume that the cutoff
$\Lambda$, a large integer, is much larger than the mass $\mu$
and the coupling
constants $g$ and $\lambda$ are small enough so that the perturbation
theory works well. As we will see, this model is ``trivial''
(see eq.(\ref{lambdaRG})), we keep the coupling constants and the
cutoff
small so that we do not approach to the Landau singularity.
The creation and annihilation operators are assumed
to satisfy the usual commutation relations.
\begin{equation}
	[a_{n},a_{m}^{\dagger}]=\delta_{n,m}.
	\label{commutator}
\end{equation}

This model is similar to the usual equal-time $\phi^{4}$ theory in
$(1+1)$ dimensions but the interaction terms which contain only
annihilation operators or only creation operators are absent. (This
is a typical feature of LFFT.)  We regard $g$ and $\lambda$ as two
independent coupling constants. Unlike the light-front $\phi^{4}$
theory, on the other hand, the ``momentum'' $n$ can take negative
values.
The unperturbed dispersion relation is also different from the usual
ones.
Therefore this $a^{4}$ theory should be regarded as a hybrid model
and
should be studied on its own right.

The reasons why we consider this model are: (1) it is extremely
simple,
and (2) it exhibits mass, coupling constant, and wave function
renormalization already at the second order in perturbation theory.
One can
study the effects of renormalization easily in this simple model.

Note that the model is invariant under the transformation
$a \leftrightarrow -a$. The eigenstates are divided into the ``odd''
sector, which contains odd number of particles and the ``even''
sector, which contains even number of particles. The Fock vacuum is
an eigenstate of the full Hamiltonian with the eigenvalue zero.
The lowest energy state above the vacuum is the (physical)
one-particle state, while the first excited state is a (physical)
two-particle state.

\section{Feynman perturbation theory}
It is easy to develop Feynman perturbation theory for the $a^{4}$
theory, though it has no covariance.
For example, the propagator is given by
\begin{eqnarray}
	G_{mn}(t) & \equiv &
	\langle 0|T a_{m}(t)a^{\dagger}_{n}(0)|0\rangle
	\nonumber  \\
	 & = & \delta_{m,n}\theta(t)e^{-i\omega_{n}t}
	\label{propagator}  \\
	 & = & \int{dE\over 2\pi}{i\delta_{m,n}\over E-\omega_{n}+
	 i\epsilon}
	 e^{-iEt}, \nonumber
\end{eqnarray}
where $a_{n}(t)$ is the operator in the interaction picture,
$a_{n}(t)=e^{iH_{0}t}a_{n}e^{-iH_{0}t}=a_{n}e^{-i\omega_{n} t}$,
where
$H_{0}=\sum_{n=-\Lambda}^{\Lambda}\omega_{n}a_{n}^{\dagger}a_{n}$
is the unperturbed Hamiltonian.
In the usual way, one can readily obtain the Feynman rules.
(See Fig.1.)

\subsection{self-energy}
In the second-order perturbation theory, the self-energy (Fig.2-a)
for the
$a^{4}$ theory is given by
\begin{eqnarray}
	-i\Sigma_{n}(E) & = & {(-ig)^{2}\over 3!}
	\sum_{p,q,r}\delta_{p+q+r,n}
	\int {dE_{p}\over 2\pi}{dE_{q}\over 2\pi}
	{i\over E_{p}-\omega_{p}+i\epsilon}
	{i\over E_{q}-\omega_{q}+i\epsilon}
	{i\over E-E_{p}-E_{q}-\omega_{r}+i\epsilon}
	\nonumber \\
	 & = & {-ig^{2}\over 3!}\sum_{p,q,r}\delta_{p+q+r,n}
	 {1\over E-\omega_{p}-\omega_{q}-\omega_{r}+i\epsilon}
	\label{self}
\end{eqnarray}
where all the momenta are cutoff at $\pm\Lambda$. One can show that
\begin{equation}
	\Sigma_{n}(E)={g^{2}\over 3!}
	\left(-3\Lambda-{3\over 2}(E-3\mu)\ln \Lambda
	+ \mbox{finite}\right).
	\label{sigma}
\end{equation}

\subsection{coupling constant}
We consider the 2-to-2 (or $\lambda$-type) interaction and
the 1-to-3 (or $g$-type) interaction separately.

For the $\lambda$-type interaction, there are two kinds of one-loop
diagrams in the second order.
The first one is a {\it fish} diagram (Fig.2-b1),
\begin{eqnarray}
	\Gamma_{fish}(m,n) & = & {(-i\lambda)^{2}\over 2}
	\sum_{p,q}\delta_{p+q,m+n}\int {dE_{p}\over 2\pi}
	{i\over E_{p}-\omega_{p}+i\epsilon}
	{i\over E_{m}+E_{n}-E_{p}-\omega_{q}+i\epsilon}
	\nonumber  \\
	 & = & {-i\lambda^{2}\over 2}\sum_{p,q}\delta_{p+q,m+n}
	 {1\over E_{m}+E_{n}-\omega_{p}-\omega_{q}+i\epsilon}
	\label{fish} \\
	& = & {-i\lambda^{2}\over 2}(-\ln \Lambda + \mbox{finite}).
	\nonumber
\end{eqnarray}
The second one is an {\it exchange} diagram (Fig.2-b2),
\begin{eqnarray}
	\Gamma_{exchange}(m,k) & = & {(-ig)^{2}\over 2}
	\sum_{p,q}\delta_{p+q,m-k}\int {dE_{p}\over 2\pi}
	{i\over E_{p}-\omega_{p}+i\epsilon}
	{i\over E_{m}-E_{k}-E_{p}-\omega_{q}+i\epsilon}
	\nonumber  \\
	 & = & {-ig^{2}\over 2}\sum_{p,q}\delta_{p+q,m-k}
	 {1\over E_{m}-E_{k}-\omega_{p}-\omega_{q}+i\epsilon}
	\label{exchange}  \\
	 & = &  {-ig^{2}\over 2}(-\ln \Lambda + \mbox{finite}). \nonumber
\end{eqnarray}
Note that there are four similar exchange diagrams.

For the $g$-type interaction, there is only one kind of one-loop
diagrams in the second order (Fig.2-c),
\begin{eqnarray}
	\Gamma_{g}(k,l) & = & {1\over 2}(-ig)(-i\lambda)
	\sum_{p,q}\delta_{p+q,k+l}\int {dE_{p}\over 2\pi}
	{i\over E_{p}-\omega_{p}+i\epsilon}
	{i\over E_{k}+E_{l}-E_{p}-\omega_{q}+i\epsilon}
	\nonumber  \\
	 & = & {-ig\lambda\over 2}\sum_{p,q}\delta_{p+q,k+l}
	 {1\over E_{k}+E_{l}-\omega_{p}-\omega_{q}+i\epsilon}
	\label{gtype}  \\
	 & = & {-ig\lambda\over 2}(-\ln \Lambda + \mbox{finite}).
	 \nonumber
\end{eqnarray}
There are three similar diagrams.

\subsection{counterterms}
In order to eliminate the divergent contributions, we add the
following counterterms,
\begin{eqnarray}
	\delta H & = &
	\sum_{n=-\Lambda}^{\Lambda}(A|n|+B)a_{n}^{\dagger}a_{n}
	\nonumber  \\
	 &  & {}+C\sum_{k,l,m,n=-\Lambda}^{\Lambda}\delta_{k,l+m+n}
	 \left(a^{\dagger}_{k}a_{l}a_{m}a_{n}
	+a^{\dagger}_{n}a^{\dagger}_{m}a^{\dagger}_{l}a_{k}\right)
	\label{counterterms}  \\
	 &  & {}+D\sum_{k,l,m,n=-\Lambda}^{\Lambda}\delta_{k+l,m+n}
	 a^{\dagger}_{k}a^{\dagger}_{l}a_{m}a_{n},
	 \nonumber
\end{eqnarray}
where
\begin{eqnarray}
	A & = & {g^{2} \over 4}\ln\Lambda,
	\label{A}  \\
	B & = & {g^{2} \over 2}\left(\Lambda-\mu\ln\Lambda\right),
	\label{B}  \\
	C & = & {g\lambda \over 4}\ln\Lambda,
	\label{C}  \\
	D & = & \left({\lambda^{2}\over 8}
	+{g^{2}\over 2}\right)\ln\Lambda.
	\label{D}
\end{eqnarray}
(We are very sloppy in fixing the finite parts of the counterterms
(the renormalization conditions).)
Note that we do not make any rescaling (wave function
renormalization) of the operators.

\section{Naive Tamm-Dancoff approximation}
In this section, we discuss a naive Tamm-Dancoff approximation and
the effects of the renormalization considered in the
previous section.
We truncate the Fock space up to including three particle
states. For
simplicity, we only consider the states with total momentum
zero. Such
a state can be expanded as
\begin{eqnarray}
	|\psi\rangle & = & c a_{0}^{\dagger}|0\rangle
	+ d{(a_{0}^{\dagger})^{2}\over \sqrt{2}}|0\rangle
	+ \sum_{n=1}^{\Lambda}
	\psi_{n}a_{n}^{\dagger}a_{-n}^{\dagger}|0\rangle
	+f{(a_{0}^{\dagger})^{3}\over \sqrt{3!}}|0\rangle
	\nonumber  \\
	 &  & {}+ \sum_{n=1}^{\Lambda}
	 \sum_{k=\left[{n+1\over 2}\right]}^{\min\{2n,\Lambda\}}
	 \left({1\over \sqrt{2}}\right)^{\delta_{k,2n}}
	 \left({1\over \sqrt{2}}\right)^{\delta_{n,2k}}
	 \varphi_{n,k}a_{n}^{\dagger}a_{-n+k}^{\dagger}a_{-k}^{\dagger}
	 |0\rangle,
	\label{state}
\end{eqnarray}
where $\left[{n+1\over 2}\right]$ stands for the largest integer
which is not larger than ${n+1\over 2}$ (Gauss' symbol).
Note that the normalization condition of the state $|\psi\rangle$ is
\begin{equation}
	\langle \psi|\psi\rangle=|c|^{2}+|d|^{2}+|f|^{2}
	+\sum_{n=1}^{\Lambda}|\psi_{n}|^{2}
	+\sum_{n=1}^{\Lambda}
	\sum_{k=\left[{n+1\over 2}\right]}^{\min\{2n,\Lambda\}}
	|\varphi_{n,k}|^{2}=1.
	\label{normalization}
\end{equation}

One can diagonalize the Hamiltonian in this restricted Fock space
numerically. Fig.~{3-a} and Fig.~{3-b} show the cutoff
dependence of the
lowest energy states
with and without the counterterms. It is easy to see that the lowest
state energy is almost independent of the cutoff if the counterterms
are included, while the first excited state energy is almost
independent if the counterterms are {\it not} included. This is
because of the TD truncation. In this restricted Fock space, one of
the particles in the two-particle sector
cannot get the self-energy
contribution, because the intermediate states would contain four
particles and are therefore out of the restricted Fock space. If one
can include, say, up to ten-particle states, the lowest energy states
would not suffer from this disease because in LFFT they usually have
negligibly small high Fock components. In practice, however,
such a large Fock
space is not feasible. It is therefore necessary to avoid this
problem.

It is important to note that the Hamiltonian itself has a very simple
structure. It can act on a state with any number of particles.
What it does is to change the particle number by two at most.
The trouble is the restriction on the {\it total} number of particles
in a {\it state}.
It suggests that we should restrict the {\it difference} of the
particle numbers in a state between before and after the interaction
is applied. The {\it difference} has no sector dependence, of course.
If we can turn off the interactions which changes the
particle number of a
state, there is then no sector dependence.
This is the basic idea of
the similarity transformation in ``particle space'' which we
discuss in the next section.

\section{Similarity transformation}
Let us first consider the similarity transformation in particle space
in a general setting. Consider a Hamiltonian $H$ of the form,
\begin{equation}
	H=H_{0}+\lambda W+ gV,
	\label{general}
\end{equation}
where $(H_{0})_{mn}=\Omega_{m}\delta_{mn}$\cite{foot2}.
The interaction $W$
does not change the particle number while $V$ does. If we includes
counterterms (through the second order in the coupling constants),
we will have additional terms (eq.(\ref{counterterms})),
\begin{equation}
	\delta H= g^{2}H_{1}+\lambda^{2}W'+g^{2}W''+ g\lambda V'+\cdots.
	\label{ct}
\end{equation}
where $W'$ and $W''$ are of $\lambda$-type and $V'$ is of
$g$-type and
$H_{1}$ which comes from the self-energy does not change
the particle
number.
We make a similarity transformation from $H$ to $H'$ by using a
unitary operator $U\equiv e^{iR}$ so that
$H'_{\Delta N\ne 0}=(UHU^{-1})_{\Delta N\ne 0}=0$, where
$H'_{\Delta N\ne 0}$
stands for the part of the Hamiltonian which changes
the particle number.
We are going to find such an operator $R$ perturbatively.
We expand $R$ in a power series of $g$ and $\lambda$,
\begin{equation}
	R=gR_{1}+g^{2}R_{2}+g\lambda T_{1}+\cdots.
	\label{R}
\end{equation}
Note that $R$ must be zero when $g=0$. Therefore $R$ does
not contain
the terms proportional only to a power of $\lambda$.

By expanding in the coupling constants,
the transformed Hamiltonian $H'$ has the form,
\begin{eqnarray}
	H'& = & H_{0} + g^{2}H_{1} +g(V+i[R_{1},H_{0}])
	+\lambda W +\lambda^{2} W''
	\nonumber \\
    & &{}+g^{2}(W''+i[R_{2},H_{0}]+i[R_{1},V]-{1\over 2}
    [R_{1},[R_{1},H_{0}]])
    \label{expanded} \\
    &  &{}+g\lambda(V'+i[T_{1},H_{0}]+i[R_{1},W]) +\cdots.
    \nonumber
\end{eqnarray}
The requirement that $H'_{\Delta N\ne 0}=0$ determines $R_{1}$,
$R_{2}$, and $T_{1}$ up to the matrix elements between the states
with the same particle number. We fix this ambiguity simply
by setting
them zero.
For example, we have
\begin{equation}
	(R_{1})_{mn}={iV_{mn}\over \Omega_{n}-\Omega_{m}},
	\label{r1}
\end{equation}
for the states $m$ and $n$ with different particle numbers
 while $(R_{1})_{mn}=0$ otherwise.
In a similar way, we have
\begin{eqnarray}
	(R_{2})_{mn} & = & {i\over \Omega_{n}-\Omega_{m}}
	\left\{W''+i[R_{1},V]
	-{1\over2}[R_{1},[R_{1},H_{0}]]\right\}_{mn},
	\label{r2}  \\
	(T_{1})_{mn} & = & {i\over \Omega_{n}-\Omega_{m}}
	\left\{V'+i[R_{1},W]\right\}_{mn},
	\label{t1}
\end{eqnarray}
for the states $m$ and $n$ with different particle numbers,
while $(R_{2})_{mn}=(T_{1})_{mn}=0$ otherwise.
The effective Hamiltonian thus takes the following form,
\begin{eqnarray}
	 H' & = & H_{0}+ g^{2}H_{1}+\lambda W+\lambda^{2}W' +g^{2}W''
	 \nonumber \\
	 &  &{}+ g^{2}\big(i[R_{1},V]
	 -{1\over2}[R_{1},[R_{1},H_{0}]]\big)_{\Delta N = 0}.
	\label{effham}
\end{eqnarray}
Note that the counterterm $V'$ does not contribute to the effective
Hamiltonian through this order. This is a good thing because the
corresponding diagrams are not included in the similarity
transformation. See Fig.4.

\section{Effective Hamiltonian}
We are now ready to obtain the effective Hamiltonian for the $a^{4}$
theory. In this section, we use the notation
\begin{equation}
	|k_{1}\cdots k_{n}\rangle\equiv {1\over
	\sqrt{n!}}\prod_{i=1}^{n}a_{k_{i}}^{\dagger}|0\rangle.
	\label{basis}
\end{equation}

\subsection{one-particle sector}
In the one-particle sector, the interactions $W$, $W'$, and $W''$ in
eq.(\ref{effham}) do not contribute.
The Hamiltonian can be written as
\begin{equation}
	H'_{k'k}\equiv \langle k'|H'|k\rangle
	=\left((1+A)|k|+\mu+B-{g^{2}\over 6}\sum_{\mathbf{l}}
	{\delta_{k,l_{1}+l_{2}+l_{3}}\over
	\sum_{i=1}^{3}\omega_{l_{i}}-\omega_{k}}\right)\delta_{k',k},
	\label{onebody}
\end{equation}
where ${\mathbf{l}}=(l_{1},l_{2},l_{3})$.
The effective interaction (the sum) comes from the
self-energy (Fig.5).
For the zero momentum state, the Schr\"odinger equation becomes
\begin{equation}
	E=\mu+B-{g^{2}\over 6}\sum^{\Lambda}_{p,q=-\Lambda \atop
		\vert p+q\vert \le \Lambda}
	{1\over |p|+|q|+|p+q|+2\mu}.
	\label{oneschroedinger}
\end{equation}

As shown in Fig.6, the eigenvalue $E$ is almost independent of the
cutoff $\Lambda$ if the counterterms are included.

\subsection{two-particle sector}
The matrix elements of the effective Hamiltonian in the two-particle
sector is
\begin{eqnarray}
	\langle k_{1}'k_{2}'|H'|k_{1}k_{2}\rangle & = & {1\over 2}\Big\{
	[(1+A)|k_{1}|+\mu+B]+[(1+A)|k_{2}|+\mu+B]
	\nonumber \\
	& & {}-{g^{2}\over
	6}\sum_{\mathbf{l}}\left[{\delta_{k_{1},l_{1}+l_{2}+l_{3}}\over
	\sum_{i=1}^{3}\omega_{l_{i}}-\omega_{k_{1}}}+
	{\delta_{k_{2},l_{1}+l_{2}+l_{3}}\over
	\sum_{i=1}^{3}\omega_{l_{i}}-\omega_{k_{2}}}\right]\Big\}
	(\delta_{k_{1}',k_{1}}\delta_{k_{2}',k_{2}}
	+\delta_{k_{1}',k_{2}}\delta_{k_{2}',k_{1}})
    \nonumber \\
	 &  & {}+{1\over 2}\Big\{ (\lambda+4D)
	 \nonumber \\
	 & &{}-{g^{2}\over 4}\sum_{p,q}\big[
	 \big({1\over \omega_{k_{1}'}+\omega_{p}
	 +\omega_{q}-\omega_{k_{2}}}
	 +{1\over \omega_{k_{1}}+\omega_{p}+\omega_{q}
	 -\omega_{k_{2}'}}\big)
	 \delta_{k_{2},k_{1}'+p+q}
	\nonumber  \\
	 &  & {}+\big(
	 {1\over \omega_{k_{1}'}+\omega_{p}+\omega_{q}-\omega_{k_{1}}}
	 +{1\over \omega_{k_{2}}+\omega_{p}+
	 \omega_{q}-\omega_{k_{2}'}}\big)
	 \delta_{k_{1},k_{1}'+p+q}
	\label{twobody}  \\
	 &  & {}+\big(
	 {1\over \omega_{k_{2}'}+\omega_{p}+\omega_{q}-\omega_{k_{2}}}
	 +{1\over \omega_{k_{1}}+\omega_{p}+
	 \omega_{q}-\omega_{k_{1}'}}\big)
	 \delta_{k_{2},k_{2}'+p+q}
	\nonumber  \\
	 &  & {}+\big(
	 {1\over \omega_{k_{2}'}+\omega_{p}+\omega_{q}-\omega_{k_{1}}}
	 +{1\over \omega_{k_{2}}+\omega_{p}+
	 \omega_{q}-\omega_{k_{1}'}}\big)
	 \delta_{k_{1},k_{2}'+p+q}\big]\Big\}
	 \delta_{k_{1}'+k_{2}',k_{1}+k_{2}}.
	\nonumber
\end{eqnarray}
It is easy to see that the terms in the first braces come from the
self-energy of two particles, while the ones in the second braces
include
the contributions from the exchange diagrams and the counterterms
(Fig.~7).

A two-particle state with zero momentum can be written as
\begin{equation}
	|\psi\rangle =\psi_{0}{1\over \sqrt{2}}
	(a_{0}^{\dagger})^{2}|0\rangle
	+\sum_{n=1}^{\Lambda}
	\psi_{n}a_{n}^{\dagger}a_{-n}^{\dagger}|0\rangle.
\end{equation}
By applying the effective Hamiltonian on this state, one can get the
Schr\"odinger equation,
\begin{equation}
	\sum_{j=0}^{\Lambda}H_{ij}\psi_{j}=E\psi_{i} \quad
	(i=0,\cdots,\Lambda)
	\label{momzero}
\end{equation}
where
\begin{eqnarray}
	H_{00} & = &
	{1\over2}\langle0|(a_{0})^{2}H'(a_{0}^{\dagger})^{2}|0\rangle
	\nonumber \\
	& = & 2\left(\mu+B-{g^{2}\over 6}\sum_{p,q=-\Lambda \atop
	|p+q|\le\Lambda}{1\over |p|+|q|+|p+q|+2\mu}\right) \nonumber \\
	& & {}-g^{2}\left({1\over 2\mu}
	+\sum_{p=1}^{\Lambda}{1\over p+\mu}\right)
	+{1\over 2}(\lambda+4D),
	\label{h00}  \\
	H_{0n} & = & H_{n0}={1\over\sqrt{2}}
	\langle0|(a_{0})^{2}H'a_{n}^{\dagger}a_{-n}^{\dagger}|0\rangle
	\nonumber \\
	& = & -{g^{2}\over \sqrt{2}}
	\sum_{p=-\Lambda+n}^{\Lambda}\left[{1\over n+|p|+|n-p|+2\mu}
	+{1\over |p|+|n-p|-n+2\mu}\right] \nonumber \\
	& & {}+{1\over \sqrt{2}}(\lambda+4D)
	\label{h0n}  \\
	H_{mn} & = &
	H_{nm}=\langle0|a_{-m}a_{m}H'
	a_{n}^{\dagger}a_{-n}^{\dagger}|0\rangle
	\nonumber \\
	& = &
	2\left[(1+A)n+\mu+B-{g^{2}\over 6}\sum_{p,q=-\Lambda
	\atop |p+q-n|\le \Lambda}^{\Lambda}{1\over
	|p|+|q|+|n-p-q|-n+2\mu}\right]\delta_{m,n}
	\nonumber \\
	& &{}-{g^{2}\over 2}\sum_{p,q=-\Lambda}^{\Lambda}
	\left({1\over m+|p|+|q|-n+2\mu}+{1\over n+|p|+|q|-m+2\mu}\right)
	\delta_{|m+p+q|,n} \nonumber\\
	& &{}+\lambda+4D,
	\label{hmn}
\end{eqnarray}
with $n,m=1, \cdots, \Lambda$. The eigenvalues can be obtained by
numerical diagonalization. The results are shown in Fig.~8.
The lowest energy state in the two-particle sector corresponds to
the first excited state of the whole Hamiltonian.
The most
important thing is that the energy of this state is almost
independent of the cutoff when the counterterms are included.
In other words, we have successfully renormalized the model so that
the TD truncation does not cause a serious problem.

\section{Discussions}
In this paper, we have proposed a new two-step renormalization
procedure and applied it to a simple model. We have obtained the
effective Hamiltonian which does not contain the
particle-number-changing interactions. The effective Hamiltonian is
diagonalized numerically. The spectrum is almost independent of the
cutoff. In this section, we discuss several aspects of our approach.

\subsection{GWW similarity renormalization}
What is the difference between the two-step approach and the
G{\l}azek-Wilson-Wegner (GWW)\cite{GW,wegner} similarity
renormalization?
Let us consider the one-particle effective interaction
for the $a^{4}$
theory induced by the GWW similarity renormalization,
\begin{equation}
	\langle k'|\Delta H|k\rangle  =  -{g^{2}\over 6}
	\delta_{k',k}\sum_{\mathbf l}\delta_{k,l_{1}+l_{2}+l_{3}}
	\theta(\Lambda-|\Delta|){1\over \Delta}
	 \theta(|\Delta|-\sigma),
	\label{GWWone}
\end{equation}
where $\Delta=|l_{1}|+|l_{2}|+|l_{3}|-|k|+2\mu$.
The sum is divergent. In the GWW similarity renormalization scheme,
the counterterms are introduced to make these contributions finite.
In this case, the counterterms to be added are
\begin{equation}
	{g^{2}\over 6}
	\left[{3\over 2}\Lambda+{3\over 2}(|k|-2\mu)\ln\Lambda\right]
	\delta_{k',k},
	\label{GWWct}
\end{equation}
up to a finite part determined by the renormalization condition.
Note that because of the different cutoff of the sum,
the counterterms
are similar to ours but different.

Now, in the standard OSU approach, one invokes ``coupling
coherence''\cite{PW}.
In the present case, it is easy to obtain the so-called coherent
Hamiltonian,
\begin{equation}
	\langle k'|H^{coh}_{\sigma}|k\rangle =
	\left\{|k|+\mu+\lim_{\Lambda\rightarrow \infty}\left[
	A|k|+B'-{g^{2}\over 6}\sum_{p,q=-\infty \atop \sigma\le
	\big|\Delta_{p,q}\big|\le \Lambda}^{\infty}
	{1\over \Delta_{p,q} }\right]\right\}\delta_{k',k},
	\label{coherent1}
\end{equation}
where $\Delta_{p,q}=|p|+|q|+|k-p-q|-|k|+2\mu$, $B'={g^{2}\over
4}(\Lambda-2\mu\ln\Lambda)$. The parameter $\sigma$ should be
considered as the renormalization scale. It is obvious that this is
essentially the same as eq.(\ref{onebody}), apart from the choice of
the cutoff.

The effective two-particle interactions induced by GWW
similarity renormalization are more complicated. To the second order,
the induced interactions obtained by lowering the cutoff
from $\Lambda$ to
$\sigma$ are
\begin{eqnarray}
	\langle k_{1}'k_{2}'|\Delta H|k_{1}k_{2}\rangle
	& = &
	-\delta_{k_{1}'+k_{2}',k_{1}+k_{2}}{\lambda^{2}\over 4}
	\sum_{l_{1},l_{2}}\delta_{l_{1}+l_{2},k_{1}+k_{2}}
	\left\{
	\theta(|\Delta^{2}|-\sigma)
	{1\over \Delta^{2}}
	\theta(|\Delta^{2}|-|\Delta^{2'}|)
	 \right.
	\nonumber  \\
	 &  &
	 {}\left.
	 +\theta(|\Delta^{2'}|-\sigma)
	{1\over \Delta^{2'}}
	\theta(|\Delta^{2'}|-|\Delta^{2}|)
	\right\}
	 \theta(\Lambda-|\Delta^{2}|)
	 \theta(\Lambda-|\Delta^{2'}|)
	\nonumber  \\
	 &  &
	 {}
	 -\left\{
	 \sum_{p,q,r}\delta_{k_{1},p+q+r}
	 \theta(|\Delta_{1}|-\sigma){1\over\Delta_{1}}
	 \theta(\Lambda-|\Delta_{1}|)
	 \right. \nonumber \\
	 & & {}\left.
	 +\sum_{p,q,r}\delta_{k_{2},p+q+r}
	 \theta(|\Delta_{2}|-\sigma){1\over\Delta_{2}}
	 \theta(\Lambda-|\Delta_{2}|)
	 \right\}(\delta_{k_{1}',k_{1}}\delta_{k_{2}',k_{2}}+
	 \delta_{k_{1}',k_{2}}\delta_{k_{2}',k_{1}}) \nonumber \\
	 & &
	 {}-\delta_{k_{1}'+k_{2}',k_{1}+k_{2}}{g^{2}\over 4}
	 \sum_{p,q}\left\{
	 \delta_{k_{1}'+p+q,k_{2}}
	 \left(
	 \theta(|\Delta_{1'2}|-\sigma){1\over \Delta_{1'2}}
	 \theta(|\Delta_{1'2}|-|\Delta_{12'}|)
	 \right.\right.
	 \nonumber \\
	 & & {}+\left.\left.
	 \theta(|\Delta_{12'}|-\sigma){1\over \Delta_{12'}}
	 \theta(|\Delta_{12'}|-|\Delta_{1'2}|)
	 \right)
	 \theta(\Lambda-|\Delta_{1'2}|)
	 \theta(\Lambda-|\Delta_{12'}|)\right.
	 \nonumber \\
	 & &{}+\delta_{k_{1}'+p+q,k_{1}}
	 \left(
	 \theta(|\Delta_{1'1}|-\sigma){1\over \Delta_{1'1}}
	 \theta(|\Delta_{1'1}|-|\Delta_{22'}|)
	 \right.
	 \label{GWWtwo} \\
	 & & {}+\left.\left.
	 \theta(|\Delta_{22'}|-\sigma){1\over \Delta_{22'}}
	 \theta(|\Delta_{22'}|-|\Delta_{1'1}|)
	 \right)
	 \theta(\Lambda-|\Delta_{1'1}|)
	 \theta(\Lambda-|\Delta_{22'}|)\right.
	 \nonumber \\
	  & &{}+\delta_{k_{2}'+p+q,k_{2}}
	 \left(
	 \theta(|\Delta_{2'2}|-\sigma){1\over \Delta_{2'2}}
	 \theta(|\Delta_{2'2}|-|\Delta_{11'}|)
	 \right.
	 \nonumber \\
	 & & {}+\left.\left.
	 \theta(|\Delta_{11'}|-\sigma){1\over \Delta_{11'}}
	 \theta(|\Delta_{11'}|-|\Delta_{2'2}|)
	 \right)
	 \theta(\Lambda-|\Delta_{11'}|)
	 \theta(\Lambda-|\Delta_{2'2}|)\right.
	 \nonumber \\
	  & &{}+\delta_{k_{2}'+p+q,k_{1}}
	 \left(
	 \theta(|\Delta_{2'1}|-\sigma){1\over \Delta_{2'1}}
	 \theta(|\Delta_{2'1}|-|\Delta_{21'}|)
	 \right.
	 \nonumber \\
	 & & {}+\left.\left.\left.
	 \theta(|\Delta_{21'}|-\sigma){1\over \Delta_{21'}}
	 \theta(|\Delta_{21'}|-|\Delta_{2'1}|)
	 \right)
	 \theta(\Lambda-|\Delta_{2'1}|)
	 \theta(\Lambda-|\Delta_{21'}|)\right.
	 \right\}
	\nonumber
\end{eqnarray}
where $\Delta^{2}=|l_{1}|+|l_{2}|-|k_{1}|-|k_{2}|$,
$\Delta^{2'}=|l_{1}|+|l_{2}|-|k_{1}'|-|k_{2}'|$,
$\Delta_{i}=|p|+|q|+|r|-|k_{i}|+2\mu$,
$\Delta_{i'j}=|p|+|q|+|k_{i}'|-|k_{j}|+2\mu$, and
$\Delta_{ij'}=|p|+|q|+|k_{i}|-|k_{j}'|+2\mu$ with $i,j=1,2$.

The counterterms necessary to make the matrix elements
eq.(\ref{GWWtwo}) finite are
\begin{eqnarray}
	\langle k_{1}'k_{2}'|\delta H_{c.t.}|k_{1}k_{2}\rangle & = &
	\delta_{k_{1}'+k_{2}',k_{1}+k_{2}}\left\{
	{\lambda^{2}\over 4}+g^{2}\right\}\ln\Lambda
	+{g^{2}\over 12}(\delta_{k_{1}',k_{1}}\delta_{k_{2}',k_{2}}
	 +\delta_{k_{1}',k_{2}}\delta_{k_{2}',k_{1}})
	\nonumber  \\
	 &  & {}
	 \times\left\{
	 \left[
	 {3\over 2}\Lambda+{3\over 2}(|k_{1}|-2\mu)\ln\Lambda
	 \right]
	 +\left[
	 {3\over 2}\Lambda+{3\over 2}(|k_{2}|-2\mu)\ln\Lambda
	 \right]
	 \right\}.
	\label{GWct2}
\end{eqnarray}

Although the structure is similar, this effective
Hamiltonian is more complicated than that of eq.(\ref{twobody}).
This is mainly because of the cutoff employed in the GWW similarity
renormalization which introduces a lot of theta functions.
(If one uses a more sophisticated smeared function, the expression
would contains integrations.) The point is that one has to examine
the divergence structure in this complicated expression in order to
obtain the counterterms.

Nevertheless, it
seems obvious that the resultant ``coherent'' Hamiltonian is
very close to that of eq.(\ref{twobody}).

The real differences between the two approaches are:
(1) There are still effective induced interactions, together with the
canonical one, which {\it change} the particle number
by two in the GWW approach (if $\sigma\ge 2\mu$). To restrict the
Fock space to the two-particle space is to throw away the
interactions.
(2) The treatment of the $\lambda$-type interactions is
different. In
the GWW approach, the counterterms for the $\lambda$-type
interactions are introduced to make the induced effective
interactions finite.
On the other hand, the counterterms in the two-step approach are
obtained in the first step altogether while the $\lambda$-type
interactions are kept intact in the second step. There is no
{\it fish} effective interaction. See Fig.~7.
(3) In the GWW approach, only the matrix elements between the
states of similar
energies survive, while in the two-step approach, any matrix
elements
between the states of the same particle number survive. This
third
point may cause a non-perturbative difference, which is discussed in
the next subsection.

The GWW similarity renormalization is elegant in the sense that it
generates an effective Hamiltonian and the counterterms in one shot.
But in practice, our two-step renormalization procedure is easier and
more systematic.

It is important to see how the ``vanishing energy denominators'' are
avoided in the two-step renormalization. The problem of ``vanishing
energy denominators'' is fake in the sense that it happens because of
our poor choice of the unperturbed states. One should instead invoke
the degenerate perturbation theory. In the Feynman perturbation
theory, on the other hand, this never becomes a problem
because in calculating Green
functions, energies (or, more appropriately, ``frequencies'') can be
kept off mass-shell.

\subsection{non-perturbative divergences}
\label{higher}
A difficult problem in the two-step approach
is the non-perturbative divergences which arise from the
$\lambda$-type interactions. In the $a^{4}$ theory, the chain
diagrams
(Fig.~{9-a}) are generated upon diagonalization of the effective
Hamiltonian in the two-particle sector. They diverge as $({1\over
2}\lambda\ln\Lambda)^{n}$, where $n$ is the number of chains. This
type of non-perturbative divergences can be canceled if
\begin{equation}
	\lambda_{\Lambda}={\lambda\over 1-{\lambda\over 2}\ln\Lambda}
	\label{lambdaRG}
\end{equation}
is used instead of $\lambda_{B}\equiv \lambda+4D$. Furthermore,
we have ladder diagrams
for the exchange interactions (Fig.~{9-b}). Upon diagonalization, a
complicated mixture of these diagrams occur. We need the
counterterms
to cancel the divergence.

Within the present approximation, the coupling constant $g$ does not
run and the one-loop renormalization group equation for $\lambda$,
\begin{equation}
	\Lambda{d\lambda_{\Lambda}\over d\Lambda}=
	{\lambda_{\Lambda}^{2}\over 2}+2g^{2},
	\label{RGE}
\end{equation}
can be solved easily. The solution is
\begin{equation}
	\lambda_{\Lambda}=
	2g\tan\left(\tan^{-1}\left({\lambda\over 2g}\right)
	+g\ln\Lambda\right).
	\label{solution}
\end{equation}
Although it improves the $\Lambda$-independence of the energies of
two-particle
states, its validity is still restricted by the perturbative nature.
(The equation (\ref{RGE}) is derived at the one-loop level.)

In the GWW approach, this kind of non-perturbative divergences does
not seem to occur because the $\lambda$-type interaction is replaced
with effective interactions which have matrix elements only between
the states of similar energies. This difference is not crucial,
however.
In the two-step approach, one may eliminate the Hamiltonian matrix
elements between the states
of very different energies by utilizing the ambiguity of $R_{mn}$
for the states $m$ and $n$ which have the same number of particles.
Alternatively, one may make a further similarity transformation for
the effective Hamiltonian in the energy space.

\acknowledgments

One of the authors (K. H.) would like to thank Robert Perry for
explaining the OSU strategy to him. He also acknowledges the
discussions with Billy Jones, Brent Allen.
Both of the authors are
grateful to Kazunori Itakura and the colleagues in Kyushu University
for the discussions.


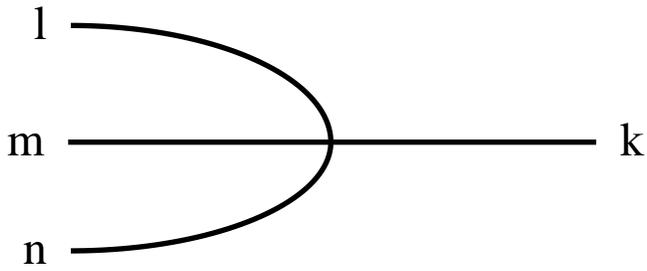
\begin{figure}
\caption{
Feynman rules for the $a^{4}$ theory.
}
\end{figure}

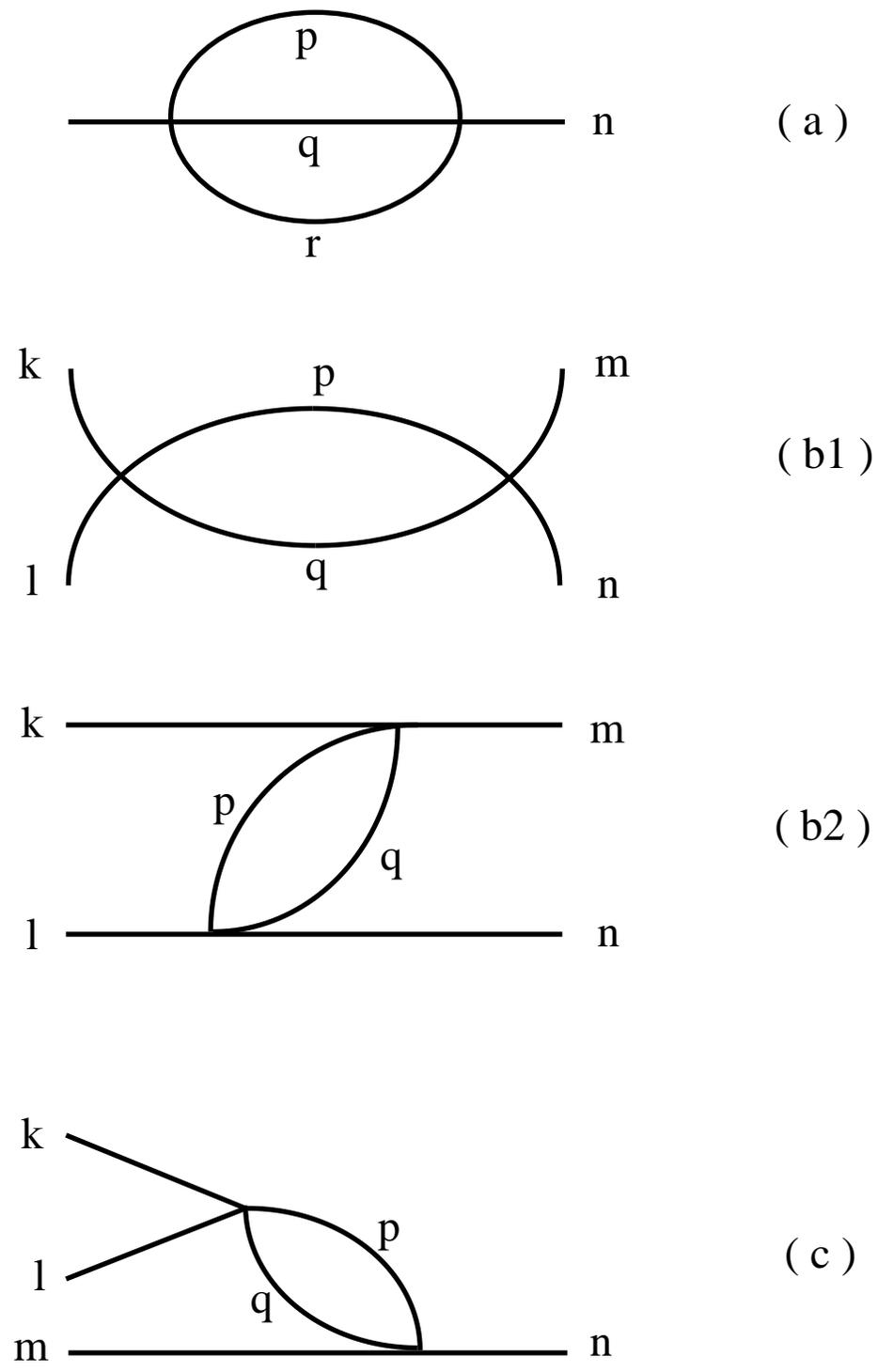
\begin{figure}
\caption{
Diagrams in the second order in perturbation theory. There are four
similar diagrams for (b2) and three for (c).
}
\end{figure}

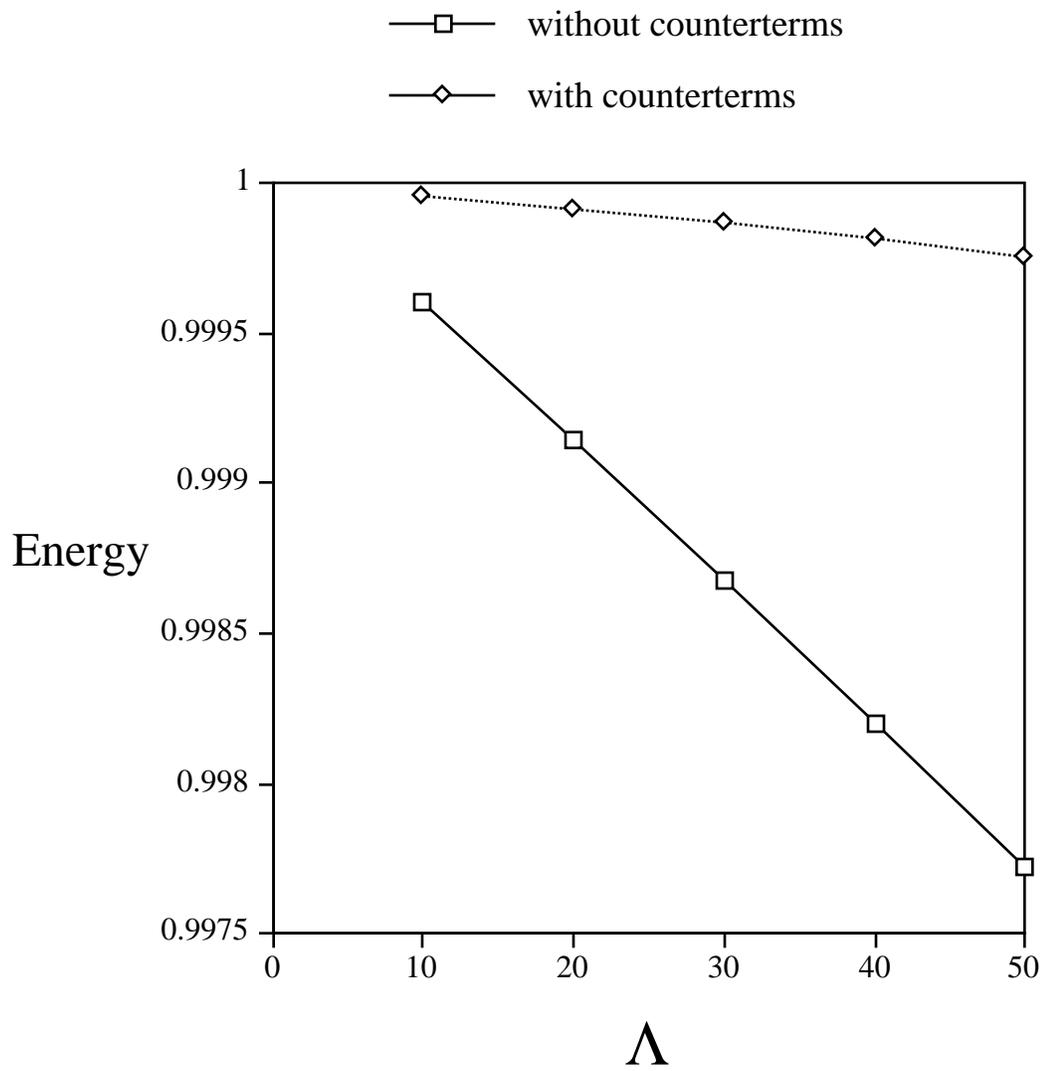
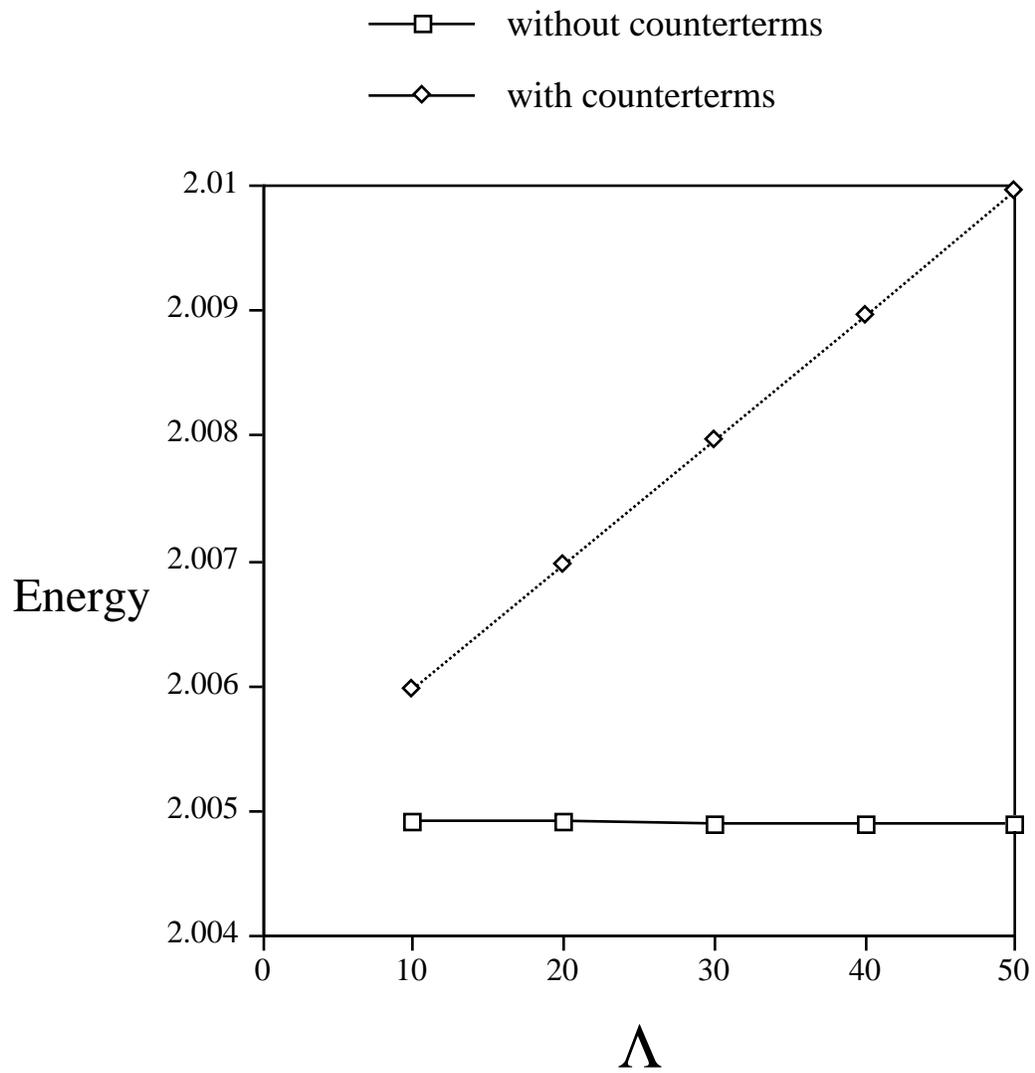
\begin{figure}
\caption{
The results in a naive TD approximation. The $\Lambda$-dependence of
the energy eigenvalue for
the lowest energy state is shown for the cases with and without
counterterms in Fig.~{3-a}.  The calculations are done for $\mu=1.0$
and $g=\lambda=0.01$. The same for the first excited state in
Fig.~{3-b}.
}
\end{figure}

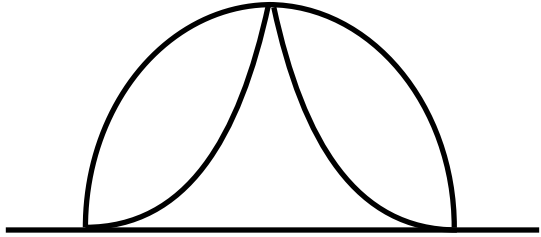
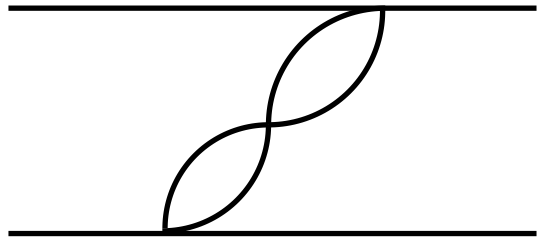
\begin{figure}
\caption{
Examples of the excluded diagrams which are relevant to the coupling
constant ($g$) renormalization.
}
\end{figure}

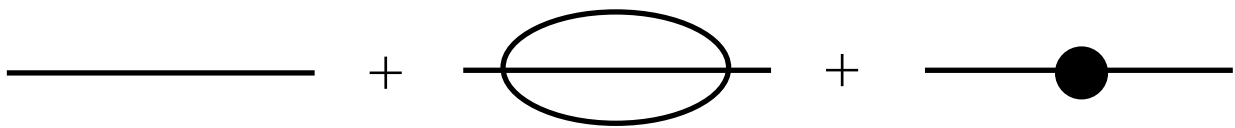
\begin{figure}
\caption{
Diagrammatic representation of the effective Hamiltonian in the
one-particle sector. The dot stands for counterterms.
}
\end{figure}

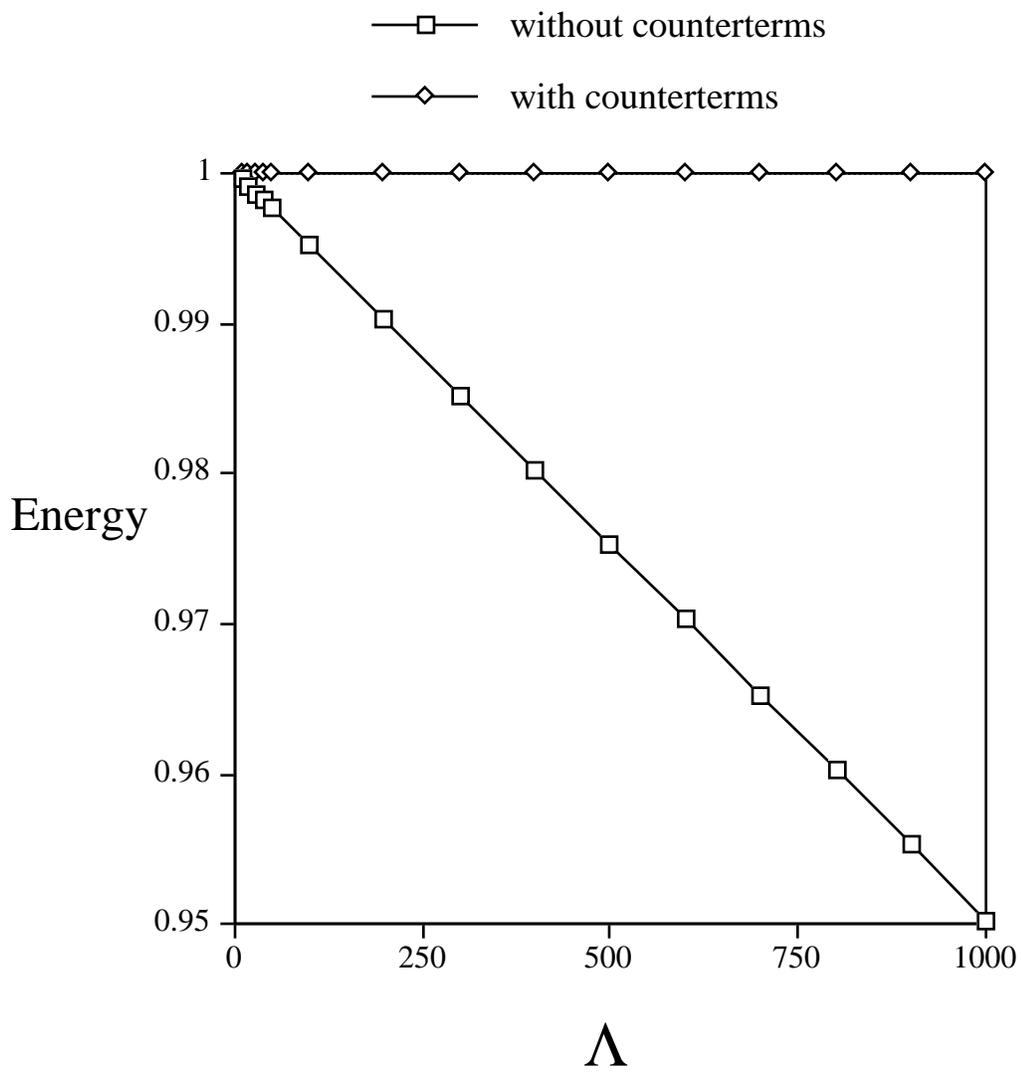
\begin{figure}
\caption{
The $\Lambda$-dependence of the energy eigenvalue for the
lowest energy
state is shown for the cases with and without counterterms.
The calculations are done for $\mu=1.0$
and $g=\lambda=0.01$.
}
\end{figure}

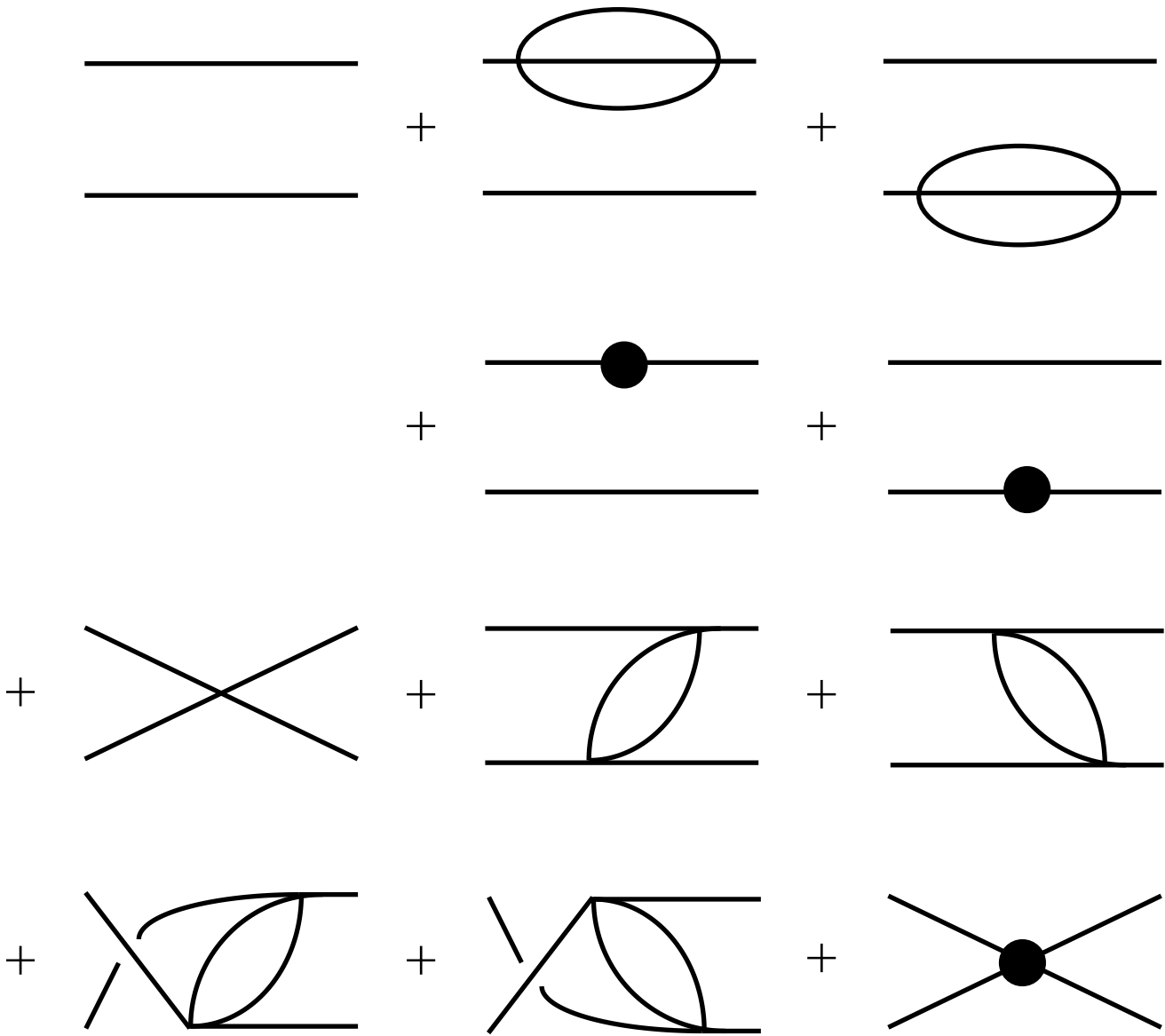
\begin{figure}
\caption{
Diagrammatic representation of the effective Hamiltonian in the
two-particle sector. The dots stand for counterterms.
}
\end{figure}

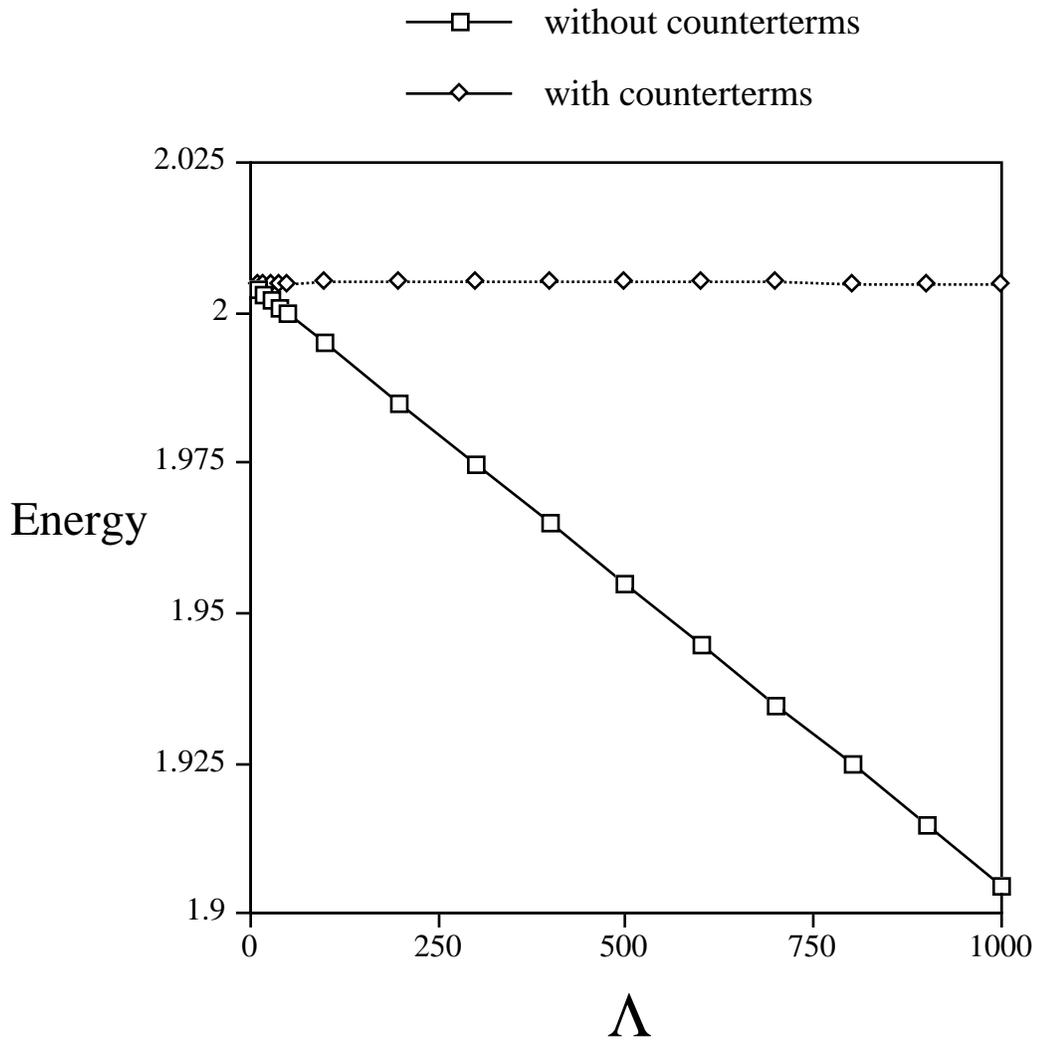
\begin{figure}
\caption{
The $\Lambda$-dependence of the energy eigenvalue for the
first excited state
is shown for the cases with and without counterterms.
The calculations are done for $\mu=1.0$
and $g=\lambda=0.01$.
}
\end{figure}

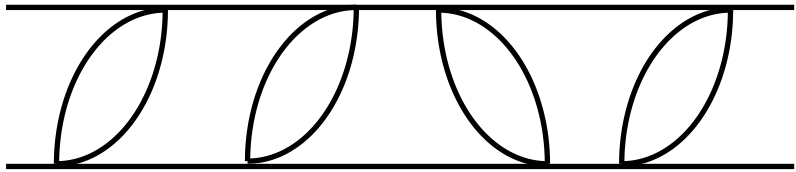
\begin{figure}
\caption{
Examples of diagrams which lead to non-perturbative divergences on
diagonalization.
}
\end{figure}


\begin{references}
\bibitem{foot}
As we will see, this second step has little to do with divergences;
they are extracted in the first step. The word ``renormalization'' is
used in its broadest meaning that the effects of the states which are
neglected are simulated by a set of effective interactions.
\end{references}
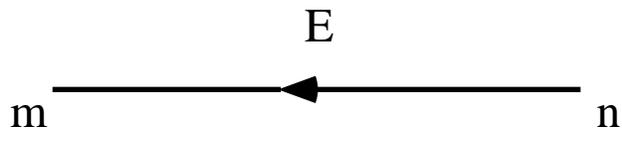
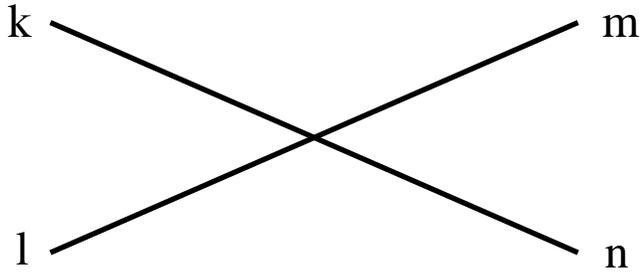
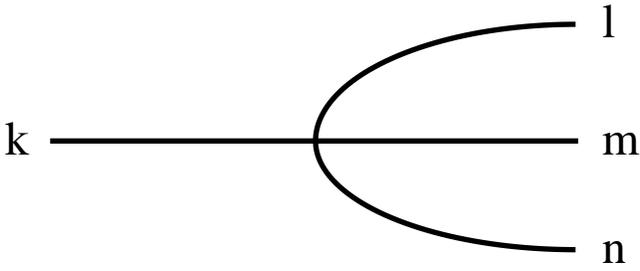
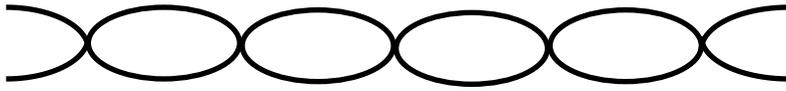
\end{document}